\documentstyle[12pt,aaspp4,epsfig]{article}
\begin{document}
\title{A molecular outflow traced by SiO masers in Orion-KL\footnote
{To appear in the Astrophyical Journal Letters}}

\author{SHEPERD S. DOELEMAN, COLIN J. LONSDALE, SHANNON PELKEY}
\affil{MIT Haystack Observatory, Off Route 40, Westford, MA 01886\footnote
{Radio Astronomy at the Haystack Observatory of the
Northeast Radio Observatory Corporation (NEROC) is supported by a grant
from the National Science Foundation}}
\authoraddr{Off Route 40\\
       Westford, MA  01886;\\
       \em dole@newton.haystack.edu, cjl@newton.haystack.edu}

\begin{abstract}
We present full resolution ($\sim0.2$mas, $\sim0.1$ A.U.) Very Long Baseline
Interferometry maps of the v=1, J=1$\rightarrow$0 SiO masers in the Orion-KL
nebula.  The morphology and turbulent nature of the maser emission argues 
against models in which the masers form in an expanding and rotating protostellar
disk.  Rather, we find that the symmetry evident in the SiO maser emission
points towards its genesis in a bipolar outflow.  Water masers which also
appear towards Orion-KL are known to share the velocity profile and double peaked characteristics
of the SiO maser emission yet form much further away from the protostar.  We link the
water masers to our present results by proposing that both species of maser form
around a flared molecular outflow of roughly constant velocity.

\end{abstract}
\newpage

\section {Introduction}

The Orion-KL nebula is a complex region of star formation whose proximity (450
pc; \cite{gen89}) provides us with a unique opportunity to study
Young Stellar Objects and their environments.  At the heart of the nebula lies
the bright infrared source IRc2 which provides most of the $\sim10^5L_\odot$
output of the region (\cite{wer83}).  Strong SiO masers are
associated with this core region which are centered not on IRc2 but have at
their centroid the nearby (0.5 as) radio continuum source ``I" (\cite{wri95,men95}).
The current interpretation is that ``I" indicates
the position of a massive young star and that IRc2 comprises illuminated
inhomogeneities in the surrounding dust envelope.  The high excitation
temperature ($\sim1800$ K) of the compact v=1 J=$1\rightarrow0$ SiO masers make 
them valuable probes of the environment close to ``I".

Based primarily on relatively low angular resolution data, a model of the Orion-KL
SiO masers has been developed which features a rotating, expanding, possibly
protoplanetary disk of inner radius $\sim40$AU and of outer radius $\sim80$AU
(\cite{bar84}).  The best structural evidence for this model
comes from spatially averaged spot maps obtained by connected element
interferometers 
(\cite{pla90,mta92,wri95,bau98}).
Such images can accurately ($\sim 0.0015\arcsec$) map
the centroids of emission at different velocities; but they reveal nothing about the
spatial structure of the masers at any given velocity on angular scales smaller
than $\sim1\arcsec$.  Early attempts at
higher-resolution study using VLBI (\cite{gre88,col92})
indicated complex structure, but did not critically test the disk model.

More recently, Schwartz et al. (1997) used the VLBA to create spot maps of SiO masers
in Orion and concluded that an alternative explanation of maser genesis in a
bipolar outflow was possible.  In this Letter we present full-resolution
($\sim$0.2mas, $\sim$0.1 A.U.) VLBA images of the Orion v=1 J=1$\rightarrow$0
SiO masers.  These images, in conjunction with published results from studies of
water masers and thermal SiO emission, allow us to discuss a model involving a
turbulent bipolar outflow oriented quite close to the line of sight, with a
flaring geometry.  In such a model, masers occur in shocked gas near the
boundaries of the flow, and the unique luminosity of the SiO masers in Orion KL
may be explained by fortuitous orientation.

\section {Observations}
We observed the Orion SiO masers ($\alpha=5^h35^m14\fs505,
\lambda=-5\arcdeg22\arcmin30\farcs45$ (J2000) (\cite{wri90})) along with
continuum calibrators on Dec 13, 1997 for a total of 10 hours.  The array
consisted of seven VLBA antennas combined with one element of the VLA.  
Full polarization
recordings of the v=1 J=1$\rightarrow$0 ($\nu_{\mbox{rest}}=43122.027$MHz)
transition were made using a
bandpasse of 8MHz, covering both the blue and red wings of the double-peaked
SiO maser spectrum.  Data were correlated at the
NRAO Socorro facility, using 512 spectral channels (15.6 kHz resolution = 0.109
km/s).  Only RR and LL products were used in forming the Stokes I images reported on here;
the full polarization data set will be presented elsewhere. 

\section {Calibration and Imaging}
The NRAO AIPS software package was used for all data reduction.  The data were
amplitude calibrated using the total-power spectra as implemented in
the AIPS task ACFIT.
This technique determines antenna gains that are contemporaneous with data collection
and corrects for antenna pointing errors.
To calibrate the instrumental delay for each antenna, a fringe fit of the
nearby continuum calibrator 0420-014 was performed.  
The resulting delay solutions for each antenna
varied by only $\sim2$ns over the entire run
resulting in estimated phase errors of no more than $\sim6\arcdeg$
accross the bandpass.  Fringe rates were determined by
fringe fitting to the brightest spectral feature in the red-shifted wing of the
cross power spectrum ($V_{\mbox{lsr}}=15.9\mbox{km}\;\mbox{s}^{-1}$).  High SNR solutions were
obtained for all stations and over the entire time range of the observations.

It became immediately apparent that in most channels,
emission was spread over a region $\sim 100$ milliarcseconds in size.  We
therefore imaged the entire field at low angular and spectral resolution
after a crude initial self-calibration, in order to locate emission suitable for
accurate self-calibration.
Fortunately, most of the compact SiO maser emission in 
Orion occurs in a handful of localized regions a few tens of milliarcseconds
across which made it possible to simultaneously image regions $\sim$150 mas apart
while accounting for all important maser features.  A suitable reference channel at
$V_{\mbox{lsr}}=12.2\mbox{km}\;\mbox{s}^{-1}$ was identified whose emission was 
contained entirely within a single sub-field.  Self-calibration solutions for this
channel were applied to the entire data set from which image cubes were then formed.

Due to space constraints, we show images of the subfields A-H in Fig. 1 that
represent the sum of all velocity channels.  These images reveal the overall
morphology of the masing region, but do not show certain low-level features
visible in individual velocity channels or groups of channels which may be
described and interpreted in the text below.

\section {Characteristics of the Maser Emission}

In accord with past and lower resolution observations, the bulk of the emission
is contained in an area of diameter $\sim$250mas ($\sim$110 AU).  There are four
main regions of emission, namely subfields A, B, (F+G) and H in Fig. 1.  The
disposition of these regions has obvious symmetry, with the center of symmetry
close to the intersection of lines joining subfields B and G, and subfields A and
H.  Subfields A and B dominate the red wing, while subfields (F+G) and H
dominate the blue wing.  A line drawn through the center of symmetry in position
angle $\sim$45 degrees cleanly separates blue-wing emission from red-wing
emission.  Below, we refer to the region north and west of this line as the ``red
zone", and the region south and east of it as the ``blue zone".

The symmetry properties of the structure extend to the substructure visible in
the various regions.  Each region has a direction defined by the presence of
long, parallel filaments of emission.  These directions do not intersect at the
overall center of symmetry, yet are roughly parallel in regions B and G, and in
regions A and H (the filaments in A are more apparent in
the individual channel maps, not displayed).  The filaments, if extrapolated,
form a ``V" in both the blue and red zones of emission whose apex lies short
of the center of symmetry, within the respective zones.

Beyond these large-scale symmetries and filaments,
the chaotic nature of both the velocity and spatial structures, and
the range of linear scales on which they occur, implies that the
masers arise in a turbulent medium.  Fig 2. shows a velocity-position contour 
image of Field G.  Integrated
intensity is shown as a function of velocity and RA across the long horizontal filaments.  A
total velocity extent of $\sim8.3$km/s characterizes the clumps and filaments in
this grouping, underscoring the turbulent nature of the masing gas.  Such
a chaotic velocity field contrasts sharply with the smooth flow assumed by
Plambeck et al. (1990) in their disk model.  The spectrum and structure derived
from a rotating and expanding disk will depend sensitively on the
lack of turbulent motion in the model.  In addition, we find no significant
compact emission present between subfields A and B, or between subfields (F+G) and H,
in contrast to the simplest interpretation of previous connected-element interferometer
data.  The velocity gradients seen in both zones by lower resolution arrays,
are the principle phenomenon which the disk model seeks to explain.
Our results indicate that the gradients are probably artifacts of 
overlapping spectra of distinct emission regions.  They may be a simple
consequence of random turbulent motions, or due to some large scale flow which
is not disk-like in nature.

We caution that the detailed disposition of the more diffuse SiO maser
emission has yet to be determined.
Only about half of the total SiO emission is represented in the images of
Fig.1., though data on our shorter baselines indicates that substantially more
emission than this is present in spatially somewhat less compact structures.
The insets show total-power spectra of the red and blue wings respectively with the
relative cross-power contributions from each sub-field.  The spectrum of the
remaining emission (shown as the dotted line in the lower right quadrant of each
inset) represents emission in features too weak or too extended to be imaged
with the array.  In both red and blue wings, these spectra are similar in shape
to the total-power spectra but are significantly smoother, as might be expected for
spatially more extended emission.  Weak compact emission is certainly present 
outside the subfields that
were imaged, but this constitutes at most a few percent of the total compact
emission, and is considered unimportant to the analysis in the next section.

\section {Discussion}

It has become clear that larger-scale molecular outflows and structures are also
centered on the continuum source ``I".  Gaume et al. (1998) show conclusively
that the `shell' water masers first isolated by Moran et al. (1977) are centered,
as are the SiO masers, on ``I" in a region roughly $2.5\arcsec$ long, highly
elongated in a PA of $\sim45\arcdeg$.  
The `shell' masers must be closely related to the
SiO masers, because the double-peaked velocity profile of the water maser lines
accurately mimics that of the SiO masers.  This important link was noted early
in the study of the Orion KL masers (\cite{sny74,mor77,gen79}).

In close proximity to ``I" are other emission features which may be dynamically
separate from the SiO masers.  On the same scales as the `shell' water masers,
SiO v=0 J=2$\rightarrow1$ emission forms a `bow-tie' oriented to the NE and SW
and centered on ``I" (\cite{wri95}).  Genzel \& Stutzki (1989) have referred
to this as a low velocity molecular flow but a flared disk morphology is also
possible.  
On larger scales, a high velocity outflow is aligned to the NW and SE.  Wright et al. (1995)
trace this component with SiO v=0 J=2$\rightarrow1$ in which they see
a weakly bipolar flow out to distances of $30\arcsec$.  In CO J=1-0, Chernin \& Wright (1996)
follow this outflow over a velocity range of $\pm50$km/s which covers the
same spatial extent.  

In light of these probably related phenomena, any interpretation of the SiO maser
structures and kinematics should relect the overall properties of
the region.  The rotating expanding disk model notably fails to integrate these
associated phenomena into a global picture and is not supported by our VLBI
results. 
This leads us to
synthesize a new model which accounts more naturally for the various
properties of the region surrounding source ``I".

We propose that the kinematics in this region is best pictured as a constant
velocity bipolar outflow whose interface with a surrounding ambient medium
produces shocks and overdensities leading to both water and SiO maser formation.
The outlines of the outflow are traced by the two main maser regions in both the
red and blue wings and the flow direction is roughly parallel to the
high velocity flow described above.  Shocks at the edges of the outflow flatten
into sheets presenting our line of sight with elongated and parallel maser
features.  Long gain paths through these shocks will mimic the tangential gain
paths traced by SiO masers in circumstellar envelopes of evolved stars (\cite{dia94,doe98}).
This explains the lack of
systematic velocity gradients along the maser filaments.  

The `shell' water masers, while sharing the SiO maser velocity profile, cover a
highly elongated NE-SW strip.  Furthermore, the velocity and spatial structure
of the water masers are not clearly ordered, with both red and blue shifted
emission occuring on both sides of the line of symmetry separating the red and
blue SiO zones.  This suggests a geometry in which the opening angle of the 
outflow flares as the flow travels from the SiO maser region
to that of the water masers resulting in a ``trumpet" shape.  A simple geometric 
requirement accounts for the relative placement of the masers and allows both maser
species to inhabit the same dynamical structure : $\theta_s < 2\theta_i < \theta_w$.
Here $\theta_s$ and $\theta_w$ are the full-width opening angles at the SiO and water 
masers respectively, with the inclination of the flow to our line of sight
represented by $\theta_i$.

In this model, projection plays a central role in determining the appearance of
the SiO and water emission.  It is likely that in a roughly constant-velocity
flow with substantial opening angle, long velocity-coherent gain paths might be
preferentially found along directions parallel to the edge of the flow.  We
speculate that the apparent luminosity of the masers in such a model would be
enhanced when the inclination of the outflow axis to the line of sight is
comparable to the opening angle of the flow.  Such an orientation produces long
gain paths in both the approaching and receding sides of the flow.
The resulting flux density enhancement  would naturally account for the fact that
the luminosity of the Orion KL masers dramatically
outstrips that of the other known SiO masers in star forming regions, in that a
fortuitous orientation renders Orion KL the only nearby member of a rare class
of aligned objects.

A thorough investigation of this model and its feasibility is beyond the scope
of this Letter.  Results of numerical simulations incorporating the above 
geometry and turbulence effects will be compared with the spatial and velocity
structures of our full resolution image cubes and presented elsewhere.

\clearpage

\section{Figure Captions}

\figcaption{Spatial and spectral structure of the Orion SiO masers.  The center plot
shows the positions of all subfields in which maser emission was detected.  The map
center is the position used during correlation and given by Wright et al. (1990).  The
surrounding grey scale images show the maser emission summed over all velocities for
each region.  The restoring beam was 0.39$\times$0.17 mas oriented in an almost purely N-S
direction.
Insets at the upper right and lower left show the 
spectral contribution of each region.  In each inset the first three quadrants show the 
total maser spectra of the red or blue wing compared with that of the subfields.  The 
last quadrant in each inset shows the sum of all red or blue wing subfields and the
difference spectra (dotted) which we label ``diffuse".  The upper left inset shows the
the single dish spectrum of the masers.  All radial velocities are relative to LSR.
\label{fig:images}
}

\figcaption{Integrated emission of the G region as a function of Right Ascension and
LSR velocity.   
The peak flux is 1.11Jy/Beam and contours are : -2, -1.4, 1.4, 2, 2.8, 4, 5.6, 8, 11.3, 16, 22.6, 32, 45.2,
 64, 90.5, 128 in units of 0.02Jy/Beam.  The line at the lower right represents a linear
distance of 1AU.} 

\end {document}
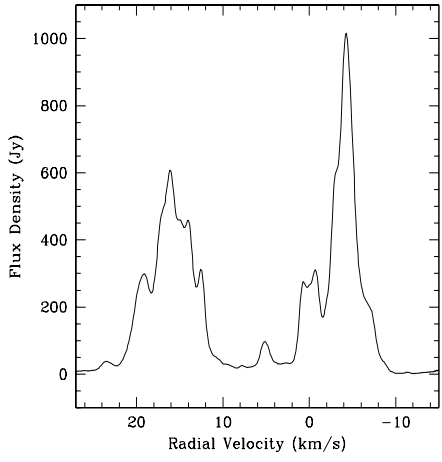
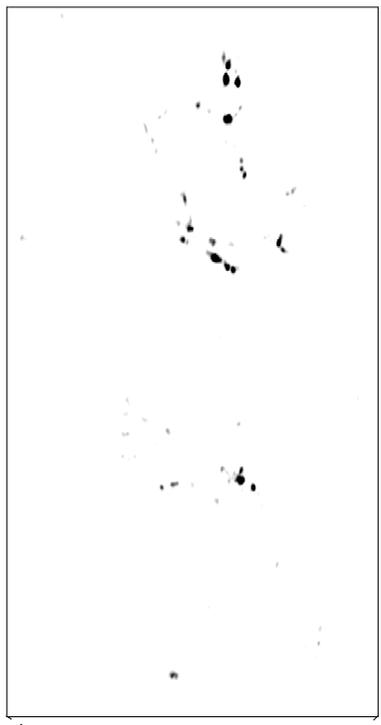
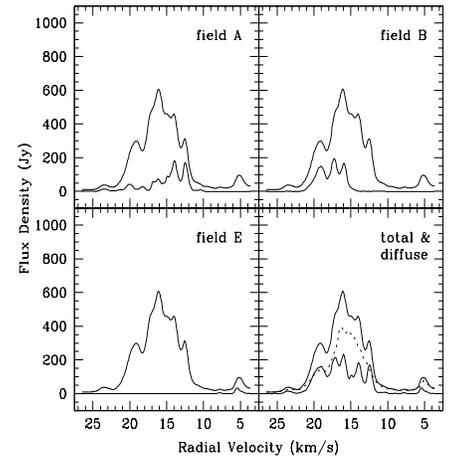
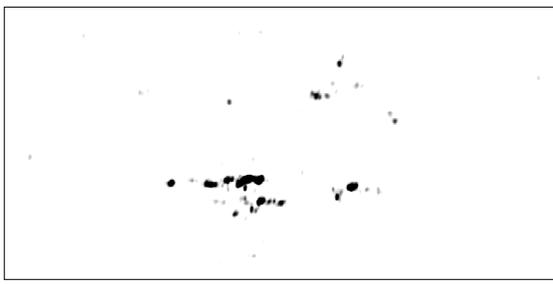
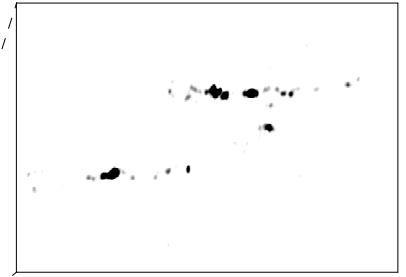
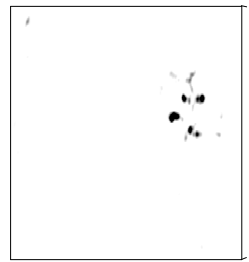
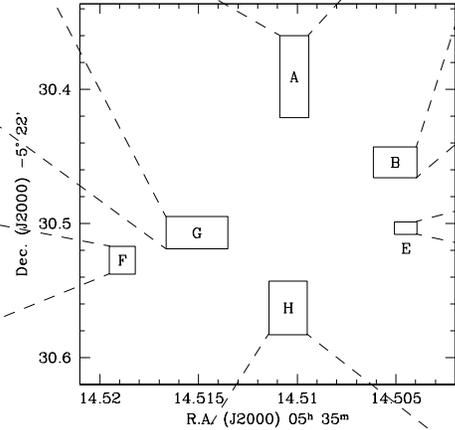
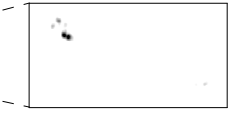
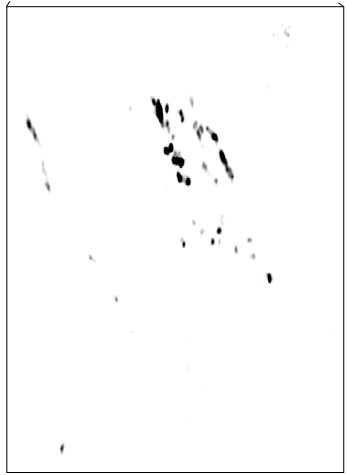
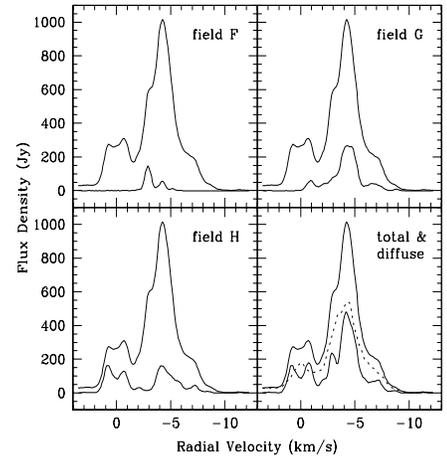
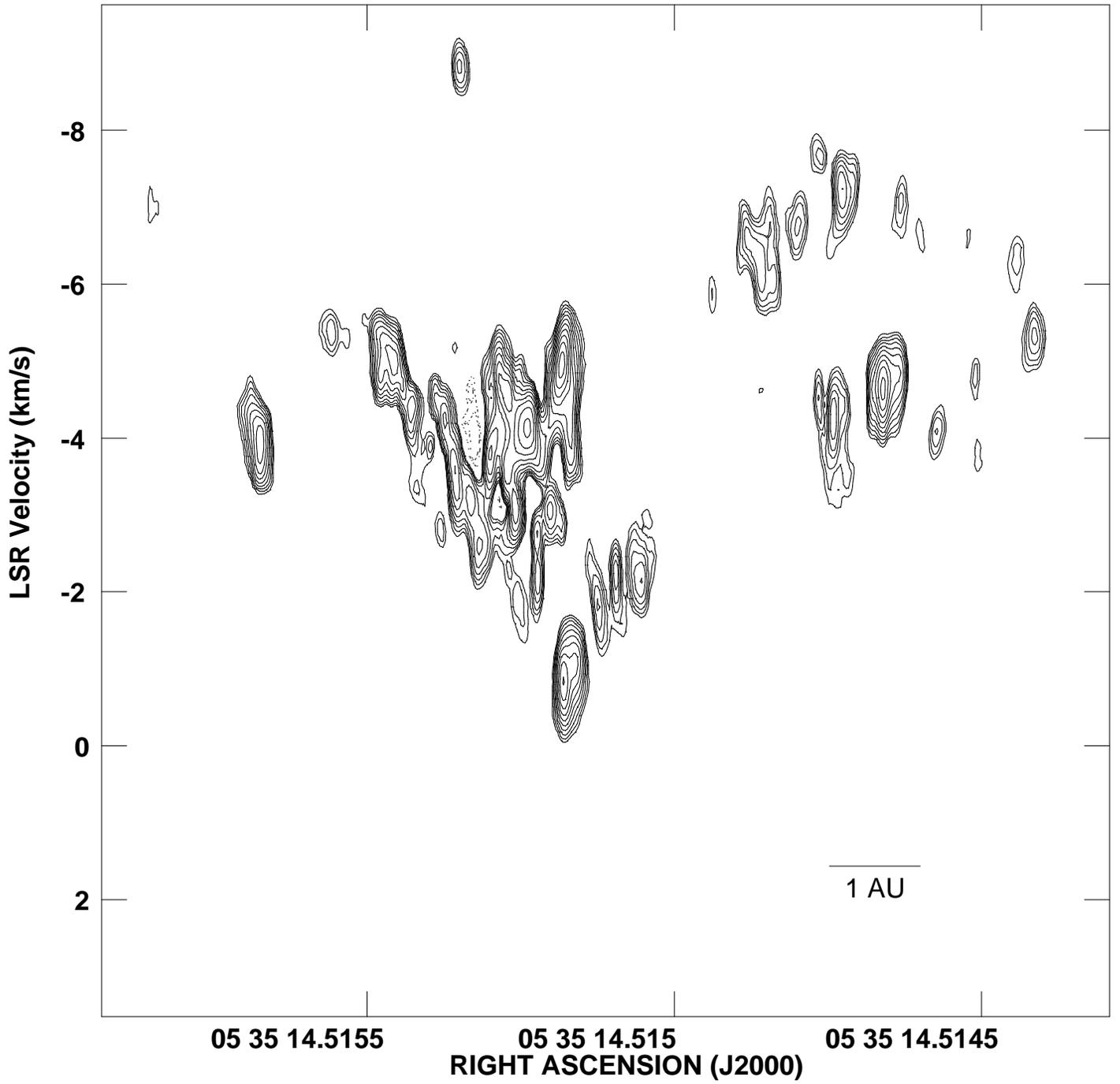
